\documentclass[12pt]{irsreport}
\usepackage{epsfig}

\title{IRS-TR 11002:  Calibration of the Acquisition Images from the Red 
Peak-Up Sub-Array}

\author{
G.C. Sloan (1) ~\& D.A. Ludovici (2) \thanks{ (1) Infrared Spectrograph
Science Center, Cornell University, (2) Department of Physics, West
Virginia University; NSF REU Research Assistant, Astronomy Department,
Cornell University} }

\date{3 August, 2011\footnote{With minor editorial corrections 23 Dec.\
2011 and corrections 24 Dec.\ 2012 to the $F_{24}$ value for HD~173511 
in Table~3 and the references.}}

\runninghead{IRS-TR 11002 - Red PU Calibration}

\begin{document}

\maketitle

\begin{abstract}

We present a calibration of the acquisition data obtained by
the Red Peak-Up (PU) sub-array on the Infrared Spectrograph
on {\it Spitzer}, based on repeated observations of three K 
giants.  This calibration is tied directly to the most 
current infrared calibration based on data from the Multiband
Imaging Photometer for {\it Spitzer}.  An analysis of the 
responsivity of the Red PU sub-array reveals no detectable 
deviations from linearity in the most recent pipeline version, 
but older pipeline versions show evidence suggesting possible 
small non-linearities.

\end{abstract}

\section{Introduction} 

One key step to the calibration of the Infrared Spectrograph
(IRS; Houck et al.\ 2004) aboard the {\it Spitzer Space
Telescope} (Werner et al.\ 2004) is the calibration of the
acquisition images from the Red Peak-Up (PU) sub-array on the
Short-Low module (SL).  Our primary standards, HR~6348, 
HD~166780, and HD~173511 were observed repeatedly during the
cryogenic {\it Spitzer} mission, and most of those 
observations began with target acquisition in the Red PU
array.  We can use these data to lock the overall photometry
from these standards, and use that to photometrically 
calibrate their spectra.

\section{Photometric calibration} 

\begin{figure} 
  \begin{center}
     \epsfig{file=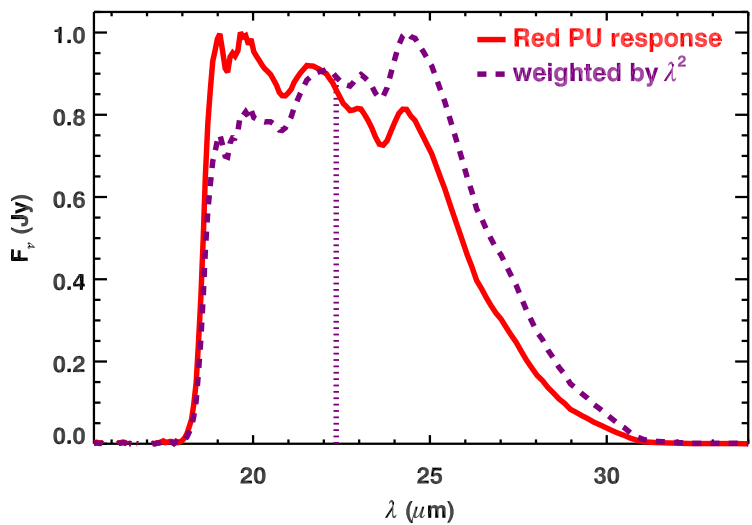, width=10cm}
  \end{center}
\caption{
---The normalized responsivity of the Red PU sub-array for a 
bias voltage of 2.0 volts, as operated during the {\it Spitzer} 
mission.  The solid curve shows the actual responsivity, while
the dashed curve shows the responsivity $\times$ $\lambda^2$.  
The centroid of this latter curve is 22.35~\mum\ (vertical 
dotted line), which we adopt as the effective wavelength of the 
Red PU filter.}
\end{figure}

To determine the effective wavelength of the Red PU filter, 
we multiply its responsivity (at a bias voltage of 2.0 volts) 
by $\lambda^2$, which approximates the long-wavelength 
behavior of a star.  The centroid of the product is 
22.35~\mum, which we will adopt for the effective wavelength 
of the Red PU sub-array.  Figure 1 illustrates the process.

We calibrate the Red PU data by tying our photometry to the
calibration of the 24-\mum\ filter on the Multiband Imaging
Photometer for {\it Spitzer} (MIPS; Engelbracht et al.\ 2007, 
Rieke et al.\ 2008).  MIPS-24 observed all three of our 
primary standards.  Our procedure is to preserve the mean of 
the MIPS-24 measurements, but to adjust the relative levels 
of the individual stars to reflect their brightnesses as 
observed in the Red PU images.

Our photometry follows the same procedure as described in 
IRS-TR~11001 (Sloan \& Ludovici 2011).  PU data were obtained 
in two sets of three images.  For each set, we produced a 
median image, then performed aperture photometry using a 
four-pixel source radius and a seven-pixel radius for the
sky annulus.  In cases where the photometry from the initial 
(acquisition) image differed from the second (sweet-spot)
image by more than 1.5\%, we rejected the initial image from
further consideration.  Data obtained after IRS Campaign 14
have been corrected for the small variations in responsivity
described in IRS-TR~1101, using divisive corrections of
0.9826 (Camp.\ 15--34) and 0.9929 (Camp.\ 35--61).

\bigskip
\noindent \begin{tabular}{lllrcccc} 
\multicolumn{5}{c}{\bf Table 1---Relative brightnesses of standards} \\
\hline
\\
{\bf Standard} & {\bf Spec.} &  & 
  \multicolumn{2}{c}{\bf Photometric signal (DN/s) } \\
{\bf star} & {\bf class} & {\bf Measurements} & 
{\bf IRS Red PU} & {\bf MIPS-24} \\
\hline
HR~6348   & K1 III & 163 & 1934.2 $\pm$ 2.1 & 30230 $\pm$ 122 \\
HD~166780 & K4 III &  79 & 1950.4 $\pm$ 4.5 & 30650 $\pm$ 884 \\
HD~173511 & K5 III & 217 & 2033.2 $\pm$ 2.0 & 31650 $\pm$ 124 \\
\hline
\\
{\bf Standard} &  &  & \multicolumn{2}{c}{{\bf Normalized signal}} \\
{\bf star}     &  &  & {\bf IRS Red PU} & {\bf MIPS-24} \\
\hline
HR~6348   & & & 0.9802 $\pm$ 0.0011 & 0.9801 $\pm$ 0.0040 \\
HD~166780 & & & 0.9870 $\pm$ 0.0023 & 0.9937 $\pm$ 0.0287 \\
HD~173511 & & & 1.0328 $\pm$ 0.0010 & 1.0262 $\pm$ 0.0040 \\
\hline
\end{tabular}
\bigskip

Table 1 presents the resulting photometry, both in digital
units from the images and normalized to unity.  Table 1 also
presents similar results from the MIPS-24 data published by
Engelbracht et al.\ (2007), using their calibration of 6.92
$\times$ 10$^{-6}$ Jy (s/DN).  The uncertainties in 
Table 1 are the uncertainties in the mean, and for the PU 
data, the lower values reflect the large number of 
(unrejected) observations.

\bigskip
\begin{tabular}{lcccc} 
\multicolumn{5}{c}{\bf Table 2---Ideal flux densities} \\
\hline
{\bf Standard} & {\bf Normalized} & {\bf Measured} & {\bf Ideal} & {\bf Ideal}\\
{\bf star} & {\bf flux density} & {\bf $F_{24}$ (mJy)} & {\bf $F_{24}$ (mJy)} &
{\bf $F_{22}$ (mJy)} \\
\hline
HR~6348   & 0.9813 $\pm$ 0.0029 & 209.2 $\pm$ 0.8 & 209.4 $\pm$ 0.6 &
  235.0 $\pm$ 0.6 \\
HD~166780 & 0.9881 $\pm$ 0.0023 & 212.1 $\pm$ 6.1 & 210.9 $\pm$ 0.5 &
  236.7 $\pm$ 0.5 \\
HD~173511 & 1.0306 $\pm$ 0.0029 & 219.0 $\pm$ 0.9 & 220.0 $\pm$ 0.6 &
  246.9 $\pm$ 0.6 \\
\hline
\end{tabular}
\bigskip

To combine the normalized signal in the 22--24~\mum\ region 
for the three sources, we took a simple average of the IRS PU 
and MIPS-24 photometry for HR~6348 and HD~173511.  For 
HD~166780, we adopted the value from the IRS PU images alone,
due to the much higher uncertainty in the MIPS data.  Table 2
presents the adopted normalized flux densities for the three
standards, shifted up 0.0011 to preserve a mean of 1.0.  The
mean of the measured MIPS-24 data is 213.4 mJy, and the 
product of this mean and our normalized flux densities is the
ideal flux density of the standards in the MIPS-24 filter.
To determine the ideal flux density in the IRS Red PU filter,
we scale the ideal MIPS-24 values by 1/$\lambda^2$, assuming
effective wavelengths of 22.35 and 23.675~\mum.

The mean ratio of the ideal 22-um flux densities to the
photometry with a four-pixel aperture gives a calibration of 
1.2151 $\pm$ 0.0008 $\times$ 10$^{-4}$ Jy (s/DN) for the IRS 
Red PU array.

\section{Linearity} 

The samples of IRS standard stars observed with the Red PU 
array and the stars with MIPS-24 photometry observed by 
Engelbracht et al.\ (2007) have ten objects in common.  
Table 3 lists their photometry with both instruments.  The
photometry for the three K giants used in the calibration 
above differs slightly because none of the data have been
rejected.  The fluxes span a factor of nearly eight, which 
allows us to investigate the linearity of the reponsivity of 
the Red PU sub-array over nearly an order of magnitude.  The 
last column of Table 3 gives the ratio of the Red PU 
photometry to the MIPS-24 photometry, after scaling the MIPS 
data by 1/$\lambda^2$.  

\begin{figure} 
  \begin{center}
     \epsfig{file=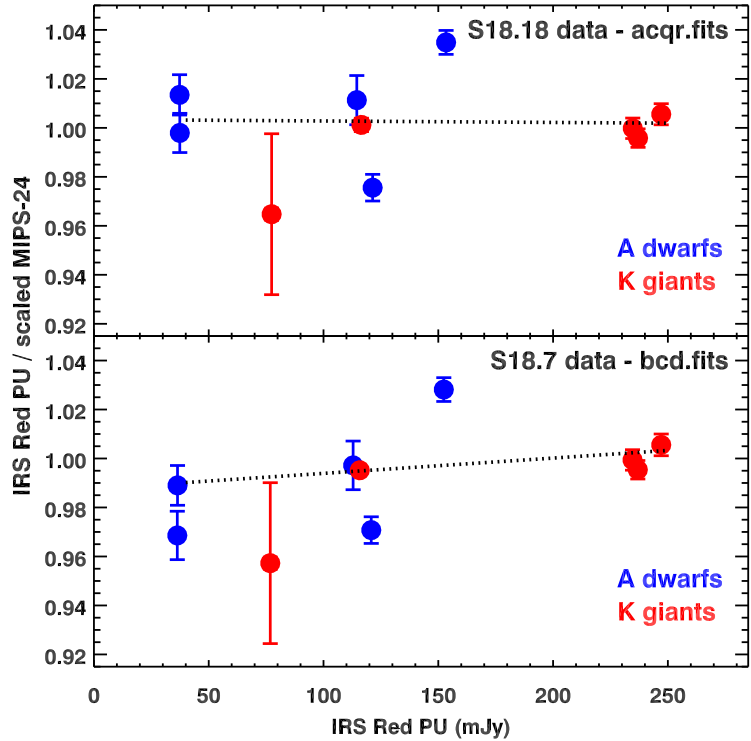, width=10cm}
  \end{center}
\caption{
---The ratio of the IRS Red PU photometry to the MIPS-24
photometry (after scaling the MIPS data to account for the
difference in wavelength), plotted vs.\ the PU flux density.
The top panel plots the data given in Table 3, which are the
S18.18 PU data, processed from the ``acqr.fits'' files.  The
bottom panel shows the S18.7 PU data, processed from the
``bcd.fits'' files.  The dotted lines are lines fitted to 
the data, accounting for the uncertainties.  No evidence for 
non-linearities in the response function are evident in the 
top panel, but the bottom panel does suggest a possible 
problem with the use of older pipeline versions of the PU 
data.}
\end{figure}

\bigskip
\begin{tabular}{lrrc} 
\multicolumn{4}{c}{\bf Table 3---Red PU and MIPS-24 photometry} \\
\hline
{\bf Standard} & {\bf IRS Red PU} & {\bf MIPS-24} & {\bf IRS Red PU /}\\
{\bf star} & {\bf $F_{\nu}$ (mJy)} & {\bf $F_{\nu}$ (mJy)} & {\bf scaled MIPS-24}\\
\hline
21~Lyn     & 114.5 $\pm$ 0.5 & 100.9 $\pm$ 0.9 & 1.011 $\pm$ 0.010 \\
26~UMa     & 153.4 $\pm$ 0.2 & 132.1 $\pm$ 0.6 & 1.035 $\pm$ 0.005 \\
HR~4138    & 121.4 $\pm$ 0.4 & 110.9 $\pm$ 0.5 & 0.976 $\pm$ 0.005 \\
HR~5467    &  37.4 $\pm$ 0.2 &  33.4 $\pm$ 0.2 & 0.999 $\pm$ 0.008 \\
HR~6348    & 234.7 $\pm$ 0.4 & 209.2 $\pm$ 0.8 & 1.000 $\pm$ 0.004 \\
HR~7018    &  37.3 $\pm$ 0.2 &  32.8 $\pm$ 0.2 & 1.012 $\pm$ 0.007 \\
HD~41371   & 116.5 $\pm$ 0.3 & 103.7 $\pm$ 0.1 & 1.002 $\pm$ 0.003 \\
HD~166780  & 237.0 $\pm$ 0.6 & 212.1 $\pm$ 0.6 & 0.996 $\pm$ 0.004 \\
HD~173511  & 247.1 $\pm$ 0.3 & 219.0 $\pm$ 1.0 & 1.006 $\pm$ 0.004 \\
BD+16~1644 &  77.4 $\pm$ 2.6 &  71.5 $\pm$ 0.4 & 0.965 $\pm$ 0.033 \\
\hline
\end{tabular}
\bigskip

Figure 2 plots the last column of Table 3 as a function of
Red PU photometry in the top panel.  The line fitted to the
data accounts for the uncertainties, and its slope deviates
from the horizontal by only 0.3~$\sigma$.  We can conclude
that the responsivity of the Red PU array behaves linearly
over the flux range considered.

The bottom panel of Figure 2 tells a different story.  Here,
the analysis is based on the older S18.7 pipeline output.  The 
S18.7 pipeline version did not provide the new ``acqr.fits'' 
files.  Instead, it provided ``bcd.fits'' files, which are 
not processed as thoroughly.\footnote{Interestingly, 
no ``bcd.fits'' files for the PU acquisition data are 
available in the S18.18 pipeline release.}  Ludovici et al.\
(2011) presented a linearity analysis of the Red PU data at
the Seattle meeting of the AAS based on the S18.7 data, and 
they concluded that in fact the data do deviate from a linear 
response, with fainter targets showing a slightly smaller 
response than brighter targets.  

Our analysis of the S18.7 data reveals a shift of $\sim$1.0\%.  
The slope deviates from the horizontal at a 3-$\sigma$ 
confidence level, which while not definitive does support the
earlier conclusion.  We conclude that the new ``acqr.fits''
files released with the S18.18 pipeline show a linear
response function, but the ``bcd.fits'' files available 
with older pipelines may not be as robust.

\end{document}